\documentstyle[preprint,aps,epsf]{revtex}
\begin{document}
\preprint{APCTP/97-11}
\draft
\title{(2+1) Dimensional Black Hole and\\ 
(1+1) Dimensional Quantum Gravity}
\author{Taejin Lee\cite{tlee}}
\address{Department of Physics, Kangwon National University, 
                      Chuncheon 200-701, Korea\\ and \\
Asia Pacific Center for Theoretical Physics, 207-43
Cheongryangri-dong\\
Dongdaemun-gu, Seoul 130-012, Korea} 

\date{\today}
\maketitle
\begin{abstract}
In the Chern-Simons gauge theory formulation of the spinning (2+1) 
dimensional black hole, we may treat the horizon and the spatial 
infinity as boundaries. We obtain the actions induced on both 
boundaries, applying the Faddeev and Shatashvili procedure.
The action induced on the boundary of the horizon is precisely 
the gauged $SL(2,R)/U(1)$ Wess-Zumino-Witten (WZW) model, 
which has been studied previously in 
connection with a Lorentz signature black hole in (1+1) dimensions. 
The action induced on the boundary of 
spatial infinity is also found to be a gauged $SL(2,R)$ WZW model, 
which is equivalent to the Liouville model, the covariant action 
for the (1+1) dimensional quantum gravity. Thus, the (2+1) 
dimensional black hole is intimately related to the quantum gravity 
in (1+1) dimensions.
\end{abstract}
%

%
The low dimensional gravity has received much attention as a laboratory
to study the interrelation between the space-time geometry and the 
quantum mechanics since the seminal works on (1+1) dimensional
gravity by Teitelboim \cite{teitelboim}, by Jackiw \cite{jackiw},
and by Polyakov \cite{polyakov87} and on (2+1) dimensional gravity 
by Deser, Jackiw, and 't Hooft \cite{deser84}, by Ach\'ucarro and
Townsend \cite{achu}, and by Witten \cite{witten88}. 
As the black hole solutions are found in (2+1) dimensional gravity
by Ba\~nados, Henneaux, Teitelboim, and Zanelli 
\cite{banados93} and in the (1+1) dimensional dilaton gravity 
by Callan, Giddings, Harvey and Strominger \cite{callan92}, it was 
understood that the difficult problems associated with the black hole
\cite{hawking75} can be dealt in low dimensions in forms far simpler 
than in (3+1) dimensions. Since then the low dimensional 
black holes have served as an important arena where one can study the 
quantum gravity, perhaps the greatest crux in theoretical physics,
in a rather tractable manner.

In the Chern-Simons formulation of the (2+1) dimensional black hole,
we may treat the horizon as well as the surface of spatial infinity 
as a  boundary. In the present paper, we show that the induced actions on 
both boundaries for the spinning (2+1) dimensional black hole 
\cite{banados93} are given by gauged $SL(2,R)$ Wess-Zumino-Witten 
(WZW) models \cite{karabali89,alekseev89}. 
However, these two gauged $SL(2,R)$
WZW models are different from each other. (We may get various
inequivalent gauged $SL(2,R)$ WZW models, depending on which subgroup is 
gauged, since the group manifold of $SL(2,R)$ has an indefinite metric.)
The induced action on the horizon is found to be equivalent to the action 
of the Lorentz signature black hole in (1+1) dimensions, which has 
been discussed extensively by Witten \cite{witten91}.
On the other hand the gauged $SL(2,R)$ WZW action induced on the surface of 
spatial infinity corresponds to the Liouville action 
which has been studied previously as the covariant action for the (1+1) 
dimensional gravity \cite{polyakov87,knizhnik88,tlee90}. It confirms the recent
result obtained by Coussaert, Henneaux and Driel \cite{couss95}. Both
cases show that the (2+1) dimensional black hole is intimately related to 
the quantum gravity in (1+1) dimensions.
 
The present paper extends the work of Carlip and Teitelboim on the
(2+1) dimensional black hole \cite{carlip951,carlip952}
and clarifies some issues: The induced boundary actions are 
{\it gauged} $SL(2,R)$ WZW models so that the difficulties 
associated with the nonunitarity of the representations of 
(ungauged) $SL(2,R)$ WZW model can be avoided.   
Since the (2+1) dimensional black hole has been 
known to be an exact solution of the effective 
action of (2+1) dimensional string 
\cite{string} and dual to a black string,
the present paper may shed some light upon these related topics also.

The BTZ (Ba\~nados-Teitelboim-Zanelli) black hole exists
\begin{eqnarray}
ds^2_{BTZ} &=& -N^2 dt^2+ N^{-2} dr^2 + r^2(N^\phi dt+ d\phi)^2,\nonumber \\
N^2(r) &=& -M +\frac{r^2}{l^2} + \frac{J^2}{4r^2}, \label{btz}\\
N^\phi(r) &=& -\frac{J}{2r^2} \nonumber
\end{eqnarray}
in the presence of a (negative) cosmological constant $\lambda=-1/l^2$.
$M$ and $J$ correspond to the mass and angular momentum of the black hole
respectively. In (2+1) dimensions the gravity is governed by a Chern-Simons 
action with an appropriate Lie-algebra valued gauge fields \cite{achu,witten88}.
In the presence of a cosmological constant the space-time is 
asymptotically anti-de Sitter, of which symmetry group is $SO(2,2)$ 
and the gravity is described by
the Chern-Simons action with the $SL(2,R)\otimes SL(2,R) \simeq SO(2,2)$ 
Lie algebra valued gauged fields
\begin{eqnarray}
I_{CS}(A,\bar A)
= \frac{k}{4\pi} \int_M \, {\rm tr} \, \left(AdA +
\frac{2}{3}AAA \right) - \frac{k}{4\pi} \int_M \, 
{\rm tr} \, \left(\bar{A}d\bar{A} + 
\frac{2}{3}\bar{A}\bar{A}\bar{A} \right) \label{csg}
\end{eqnarray}
where $k= -\frac{l}{4G},  \quad A = A^{a}J_a, \quad 
\bar A = \bar A^{a}\bar J_a$ and
\begin{eqnarray}
\left[ J_a , J_b \right] &=& {\epsilon_{ab}}^c J_c, \qquad
\left[ \bar J_a , \bar J_b\right] = {\epsilon_{ab}}^c \bar J_c, \qquad
\left[ J_a , \bar J_b \right] = 0. \nonumber
\end{eqnarray} 
Here $G$ is the gravitational constant.
The equation of motion for the gauge field implies that
the gauge field is a pure gauge and may be written in terms of multivalued 
gauge functions $u$ and $\bar u$ as 
$A= u^{-1}d u$, $\bar A= \bar u^{-1}d \bar u$ \cite{cangemi92}.

The BTZ black hole solution Eq.(\ref{btz}) has two horizons:
outer one at $r = r_+$ and inner one at $r = r_-$
where $r_\pm$ are two zeros of $N(r)$.
Thus, one may divide the space into
three regions bounded by the horizons: $0< r < r_{-}$ ($\Sigma_I$), 
$r_{-}< r < r_{+}$ ($\Sigma_{II}$), $r_{+}< r < \infty$ ($\Sigma_{III}$).
Since the curvature is constant everywhere, it seems ad hoc
to divide the space such a way. However, if we are concerned
with the quantum theory using the path integral, it seems 
unavoidable to confine ourselves to the space-time
$M=\Sigma_{III} \times R$. $M$ has two boundaries, $\partial M_1$ at 
$r=r_+$ and $\partial M_2$ at the spatial infinity.

As is discussed in ref. \cite{elitzur89}, we need to supplement the 
Chern-Simons gravity action Eq.(\ref{csg}) by boundary terms
so that the boundary conditions are imposed consistently.
For each boundary, an appropriate boundary term will be introduced.
Since we can discuss both boundary actions in a similar way, we will
proceed with the boundary action on $\partial M_1$ first.
It is convenient to introduce a new coordinates $(\tau, \rho, \phi)$ 
to discuss the BTZ black hole in the region $M$, where $\tau = t/l$ and
$r^2 = r^2_+ \cosh^2\rho - r^2_- \sinh^2\rho$. In terms of new coordinates,
we see that the BTZ black hole solution has a chiral structure:
\begin{mathletters}
\label{class:all}
\begin{eqnarray}
&&\left ( \begin{array}{l}
 A_R = \frac{2}{l}(r_+-r_-)
 \left(\sinh\rho J_0 +\cosh\rho J_2\right), \\
 A_L = 0, \\
 A_\rho = J_1, \\
\end{array} \right. \\
&&\left ( \begin{array}{l}
 \bar A_R = 0 \\
 \bar A_L = -\frac{2}{l}(r_++r_-)
 \left(\sinh\rho \bar J_0 -\cosh\rho \bar J_2\right), \\
 \bar A_\rho = - \bar J_1. 
\end{array} \right.
\end{eqnarray}
\end{mathletters}
where $A_{R/L} = A_\tau \pm A_\phi$, and 
$\bar A_{R/L} = \bar A_\tau \pm \bar A_\phi$.  
The boundary values of the Chern-Simons gauge fields on 
$\partial M_1$ read from the above solution are
\begin{eqnarray}
A_R &=& \frac{2}{l} (r_+-r_-) J_2, \quad A_L = 0, \quad A_\rho = J_1, \\
\bar A_R &=& 0, \quad \bar A_L = \frac{2}{l}(r_++r_-) \bar J_2, \quad
\bar A_\rho = - \bar J_1. \nonumber
\end{eqnarray}
These boundary values suggest us to take the boundary conditions as
\begin{equation}
A_L = 0, \quad \bar A_R = 0. \label{bc}
\end{equation}
(With other choices, the induced actions on the boundaries
may suffer unitarity problem.)
In order to impose the boundary condition consistently, 
we should introduce the boundary term as follows
\begin{eqnarray}
I_B &=& -\frac{k}{4\pi} \int_{\partial M}\, 
{\rm tr} (A_\tau - A_\phi) A_\phi + \frac{k}{4\pi} \int_{\partial M}\, 
{\rm tr} (\bar A_\tau + \bar A_\phi) \bar A_\phi, \label{bdaction}
\end{eqnarray}
where $\partial M =\partial M_1$.
This boundary term is chosen such that its variation cancels that of 
the Chern-Simons action.

The gauge invariance of the action is now broken,
partly because the space-time $M$ has boundaries, and partly
because the boundary terms do not respect the gauge invariance. 
As a consequence, the degrees of freedom of the gauge fields 
corresponding to the broken symmetry cannot be gauged away 
any longer. They become dynamical degrees of freedom as Carlip 
discussed \cite{carlip951}. To get the proper action for 
these degrees of freedom, we resort to the Faddeev and 
Shatashvili (FS) proposal for the consistent quantization 
of anomalous theory \cite{faddeev86}.

The FS proposal is to introduce a one-cocycle in such a way
the local gauge symmetry is restored and to use it as the action
describing the ``would be" gauge degrees of freedom. This
procedure has been applied to construct the action for
the (1+1) dimensional quantum gravity \cite{tlee90}.
The one-cocycle for the Chern-Simons gravity is constructed to be
\begin{eqnarray}
\alpha_{G}[A,\bar A,g,\bar g] 
&=& I_{CS}(A^g,\bar A^{\bar g})+I_B(A^g,\bar A^{\bar g}) 
- I_{CS}(A,\bar A)-I_B(A,\bar A)\\
A^g &=& g^{-1} d g + g^{-1} A g \nonumber \\
\bar A^{\bar g} &=& {\bar g}^{-1} d {\bar g} + 
   {\bar g}^{-1} {\bar A} {\bar g}. \nonumber
\end{eqnarray} 
One sees that $\alpha_{G}[A,\bar A,g,\bar g]$ satisfies the
one-cocycle condition by construction as usual 
\begin{eqnarray}
\delta \alpha_{G} &=& \alpha_{G}[A^h, \bar A^{\bar h}, g,\bar g] - 
\alpha_{G}[A,\bar A, hg, {\bar h}{\bar g}] 
+\alpha_{G}[A, \bar A, h,\bar h]=0
\end{eqnarray} 
and thanks to it, the gauge symmetry $SL(2,R) \otimes SL(2,R)$ is 
fully restored.
The explicit expression for the one-cocycle is
\begin{eqnarray}
\alpha_{G}(A, \bar{A},g,{\bar g})
&=& \alpha_1(A,g)+ {\bar \alpha}_1(\bar{A},{\bar g}), \label{acsg} \\
\alpha_1(A,g) &=& \Gamma^L[g] +
\frac{k}{2\pi}\int_{\partial M} {\rm tr}
(\partial_\phi g g^{-1})A_{L}, \nonumber\\
\bar \alpha_1(\bar A,\bar g) &=& - \Gamma^R[\bar g] - 
\frac{k}{2\pi}\int_{\partial M}{\rm tr} (\partial_\phi 
{\bar g}{\bar g}^{-1}){\bar A}_{R}, \nonumber \\
\Gamma^L[g] &=& \frac{k}{4\pi} \int_{\partial M} {\rm tr}
(g^{-1}\partial_- g)(g^{-1} \partial_\phi g) -\frac{k}{12\pi}
\int_{M} {\rm tr} (g^{-1} dg)^3, \nonumber\\
\Gamma^R[{\bar g}] &=& \frac{k}{4\pi} \int_{\partial M} {\rm tr}
({\bar g}^{-1}\partial_+ {\bar g})({\bar g}^{-1} \partial_\phi {\bar g}) 
- \frac{k}{12\pi}\int_{M} {\rm tr} ({\bar g}^{-1} 
d{\bar g})^3 \nonumber
\end{eqnarray}
where $\partial_{\pm} = \partial_\tau \pm \partial_\phi$. 
If $\partial M$ is the boundary of the spatial infinity, 
one should replace $k$ by $-k$ in the action except for the 
coefficients of the WZ terms. 
Eq.(\ref{acsg}) shows that the induced action on $\partial M_1$
is given by a direct sum of two chiral WZW actions: 
one with left moving chiral boson field only
and the other with right moving chiral boson field only. The second terms
in $\alpha_1(A,g)$ and $\bar \alpha_1(\bar A,\bar g)$ describe 
coupling of the chiral bosons to the gauge fields. (These terms
will be important when we evaluate the black hole entropy.) 
Due to the Gauss' constraint the gauge fields do not have
local degrees of freedom. Once a gauge condition
and boundary conditions are chosen appropriately, the gauge 
fields would be given uniquely by the classical BTZ black hole solution.

Applying the FS proposal to the BTZ black hole system, we find that
the quantum action induced on the boundary is described by
two chiral WZW models with opposite chirality, which
are coupled to the classical BTZ black hole background. 
Making use of the Polyakov-Wiegman identity, these two chiral 
fields can be interpreted as left and right moving modes of 
a non-chiral WZW model \cite{couss95}
\begin{equation}
I_1 = \Gamma^L[g]+ \Gamma^R[{\bar g}^{-1}] = \Gamma[{\bar g}^{-1} g] 
\equiv \Gamma[h].
\end{equation}
Therefore, we may conclude that the quantum induced action is given as the
non-chiral $SL(2,R)$ WZW model.
However, we must note that the gauge symmetry, 
$SL(2,R)\otimes SL(2,R) \simeq SO(2,2)$ 
is not completely broken by the classical BTZ black hole solution. 

Recall that the three dimensional geometry is completely determined by 
holonomies or Wilson loops of the Chern-Simons gauge fields \cite{holo}
\begin{eqnarray}
W[C] &=& {\cal P} \exp \left( \oint_C A_\mu dx^\mu \right), \\
\bar W[C] &=& {\cal P} \exp \left( \oint_C \bar A_\mu dx^\mu \right)\nonumber
\end{eqnarray}
and the holonomies depend only on the homotopy class of $C$, 
where $C$ is a closed curve and ${\cal P}$ denotes a path ordered product.
We observe that the holonomies do not depend
on the starting point of the curve $C$ on $\partial M_1$.
In the BTZ black hole the only homotopically nontrivial
closed curve is $C$: $\phi(s)= 2\pi s$,
$s\in [0,1]$ and all other homotopically
nontrivial ones can be given as products of $C$'s. 
The holonomies transform under gauge transformation as
\begin{eqnarray}
W[C] \rightarrow gW[C]g^{-1}, \quad 
\bar W[C] \rightarrow {\bar g} W[C]{\bar g}^{-1}.
\end{eqnarray}
If we take $C$ a space-like closed curve on $\partial M_1$,
\begin{eqnarray}
W[C] &=& \exp \left[ \frac{2\pi}{l}(r_+-r_-)J_2 \right], \\
\bar W[C] &=& \exp \left[-\frac{2\pi}{l}(r_++r_-) \bar J_2 \right]. \nonumber
\end{eqnarray}

Considering the following gauge transformation generated by
\begin{equation}
\Lambda = \exp(f J_2), \quad
{\bar \Lambda} = \exp({\bar f} {\bar J_2})
\end{equation}
we find that the holonomies are invariant under the gauge transformation
generated by $\Lambda$ and ${\bar \Lambda}$. 
It implies that we should equally take 
$A_{cl}^{\Lambda} = \Lambda^{-1} d\Lambda+ \Lambda^{-1} A_{cl}\Lambda$
(${\bar A}_{cl}^{{\bar \Lambda}} = {\bar \Lambda}^{-1} 
d{\bar \Lambda}+ {\bar \Lambda}^{-1} {\bar A}_{cl}{\bar \Lambda}$)
as the classical background on the boundary as well as $A_{cl}$ 
(${\bar A}_{cl}$) given by Eq.(\ref{class:all}).
Taking this into account we may write the
path integral representing the generating functional as
\begin{eqnarray}
Z &=& \int D[\Lambda, {\bar \Lambda}] D[g,{\bar g}]
\exp\left\{iI_G(A^\Lambda,\bar{A}^{\bar \Lambda},g,\bar{g})\right\} \\
&=& \int D[\Lambda, {\bar \Lambda}] D[g,{\bar g}]
\exp\left\{iI_G(A,\bar{A},\Lambda^{-1}g,{\bar \Lambda}^{-1}\bar{g})
\right\}, \nonumber \\
I_G &=& I_{CS}(A,{\bar A})+I_{B}(A,{\bar A})
+\alpha_G(A,{\bar A},g,{\bar g}). \nonumber
\end{eqnarray}
Here we have chosen appropriate gauge fixing conditions for
the gauge fields. 
Hence, the resultant action is a $U(1)$ gauged WZW model on $\partial M$
and this $U(1)$ gauge group should not be taken
into account in construction of $\alpha_{G}[A,\bar A,g,\bar g]$,
which is introduced to restore the gauge invariance. That is,
the $U(1)$ subgroup generated by $J_2$ should be gauged. Then the 
correct induced quantum action must be a $SL(2,R)/U(1)$ WZW model
\begin{eqnarray}
I_1 &=& \Gamma[h] + \frac{k}{4\pi} \int_{\partial M} 
{\rm tr} \bigl(B_+ \partial_- h h^{-1} - B_- h^{-1}\partial_+ h 
+B_+ h B_- h^{-1} - B_+B_-\bigr)
\end{eqnarray} 
where $B_\pm = B_\pm^2 J_2$.
This gauged WZW action has been found to depict a Lorentzian black hole
in (1+1) dimensions by Witten \cite{witten91}: The target manifold
of the gauged WZW model can be understood as a (1+1) dimensional
black hole with a Lorentz signature \cite{comm1}. 

Construction of the induced action on the boundary at spatial infinity
can be proceed in parallel to that of the induced action on the horizon.
As $r \rightarrow \infty$ (equivalently, $\rho \rightarrow \infty$),
\begin{eqnarray}
A_R &\rightarrow & \frac{e^\rho}{l} (r_+-r_-) 
J_+, \quad A_L = 0, \quad A_\rho = J_1, \label{bc2} \\
\bar A_R &=& 0, \quad 
\bar A_L = - \frac{e^\rho}{l}(r_++r_-) \bar J_-, \quad
\bar A_\rho = - \bar J_1 \nonumber
\end{eqnarray}
where $J_\pm = J_0 \pm J_2$.
In order to impose the boundary conditions, same as Eq.(\ref{bc}) on
$\partial   M_2$,  we   should  also   introduce  the   boundary   terms  on   $\partial  M_2$    
(Eq.(\ref{bdaction})). Following the same procedure we applied to the
induced action on $\partial M_1$, we find that the induced action on
$\partial M_2$ can be given also by the $SL(2,R)$ WZW model. However,
we should take note of difference in the boundary values. The
boundary values of the gauge fields, Eq.(\ref{bc2}) have a structure 
different from that of the boundary values on $\partial M_1$ so 
that the subgroup to be gauged is not the same. 
As a result, we will get a different
gauged WZW model as the induced quatum action on $\partial M_2$.

For a closed curve $C$ on $\partial M_2$ the Wilson 
loop elements are
\begin{eqnarray}
W[C] &=& \left(\begin{array}{cc} 1 & 0 \\ 
\frac{e^\rho\pi}{l} (r_+-r_-) & 1 \end{array} \right), \quad \\
\bar W[C] &=& \left(\begin{array}{cc} 1 & 
\frac{e^\rho\pi}{l}(r_++r_-)\nonumber \\
0 & 1 \end{array} \right).
\end{eqnarray}
It follows that gauge group elements that leave the Wilson loop element 
invariant have the following forms:
\begin{eqnarray}
g =  \left(\begin{array}{cc} 1 & 0 \\ f & 1 \end{array} \right), \quad
\bar g =  \left(\begin{array}{cc} 1 & {\bar f} \\ 0 & 1 \end{array} \right).
\end{eqnarray}
These subgroups are unbroken on the boundary of spatial infinity and
should be factored out in construction of the induced quantum action as before.
But in order to gauge these subgroups we need to take some care. 
When these subgroups are gauged, the WZW model takes the form different
from that of the WZW model for the coset conformal field theory \cite{karabali89}, 
which is the case for the quantum action induced on $\partial M_1$.
Gauging these subgroups can be taken care of by introducing constraints
as follows
\begin{eqnarray}
I_2 &=& \Gamma[h] + \int_{\partial M_2} {\rm tr} \Bigl \{\lambda_1 
\left(h^{-1} \partial_+ h J_+ -a\right)
+ \lambda_2 \left( \partial_- h h^{-1} J_- -b\right) \Bigr\}
\end{eqnarray}
where $a$ and $b$ are some appropriate constants.
Here $\Gamma[h]$ is defined on  $\partial M_2$.
These constraints, which can be rewritten in terms 
of $g$ and ${\bar g}$
\begin{eqnarray}
{\rm tr} \Bigl\{ \lambda_1 
\left(g^{-1} \partial_+ g J_+ -a\right)+ \lambda_2 
\left( \partial_- {\bar g}^{-1} {\bar g} {\bar J}_- -b\right)\Bigr\}
\end{eqnarray}
generate the gauge transformation which leaves the Wilson loop elements invariant.
As is well known in the study of covariant action for two dimensional gravity,
this   constrained   $SL(2,R)$   WZW    model   is   equivalent   to   the   Liouville   model 
\cite{alekseev89,tlee90}.
Making use of the Gauss decomposition $h= ABC\omega$,
\begin{eqnarray}
A &=& \left(\begin{array}{cc} 1 & v \\ 0 & 1 \end{array} \right), \quad
B = \left(\begin{array}{cc} e^{\varphi/2} & 0 \\ 0 & 
e^{-\varphi/2} \end{array} \right), \\
C &=& \left(\begin{array}{cc} 1 & 0 \\ w & 1 \end{array} \right), \quad
\omega = \pm \left(\begin{array}{cc} 0 & 1 \\ -1 & 0 \end{array} \right), \nonumber
\end{eqnarray}
and integrating out the Lagrangian multipliers $\lambda_1$ and $\lambda_2$,
we get the Liouville model
\begin{eqnarray}
I_2 &=& \frac{1}{8\pi} \int_{\partial M_2} \left(\partial_+\varphi \partial_-\varphi
+ (k ab) \exp(\beta\varphi)\right),\\
\beta &=& \sqrt{\frac{2}{|k|}}. \nonumber
\end{eqnarray}
Thus, we confirm the result of Coussaert, Henneaux and Driel obtained in their
study on the (2+1) dimensional black hole \cite{couss95}.

The present paper will be concluded with a few remarks.
We obtained the induced actions for the spinning (2+1)
dimensional black hole on the boundary which consists of the horizon and the surface
of spatial infinity, adopting the Faddeev-Shatashvili procedure,
which yields gauge invariant actions. Resultant induced quantum actions
are indentified as $SL(2,R)$ WZW models for both cases. 
However, we point out that some subgroups of $SL(2,R)$ are unbroken
so that the corresponding degrees of freedom should not be taken into 
account as physical ones. Thus, the correct induced quantum actions
are gauged WZW models. For the induced action on the horizon
we obtain the $SL(2,R)/U(1)$ WZW model and for that on the spatial
infinity, the Liouville model. It is interesting to note
that the two dimensional quantum gravity is essential to understand
the (2+1) dimensional black hole: the $SL(2,R)/U(1)$ WZW model describes 
a Lorentz black hole in (1+1) dimensions and the Liouville
model can serve as a covariant action for the (1+1) dimensional
quantum gravity. Since both induced actions are known exactly soluble, 
the present paper supports the work of Witten \cite{witten88}, 
which asserts that the (2+1) dimensional gravity is exactly soluble 
at the classical and quantum levels. In contrast to the (ungauged)
$SL(2,R)$ WZW model, the gauged $SL(2,R)$ WZW models we discussed
as the induced quantum actions have unitary representations.
So the entropy of the BTZ black hole may be evaluated in a precise
manner and the works of Carlip and Teitelboim \cite{carlip951,carlip952}
can be improved. We also discussed the asymptotic dynamics of the
BTZ black hole and confirmed the result of ref.\cite{couss95}
in the same framework. The present paper shows clearly that
the (2+1) dimensional black hole is intimately related to the 
quantum gravity in (1+1) dimensions.

This work was supported in part by KOSEF Research Project 951-0207-052-2
and by the Basic Science Research Institute Program, Ministry of
Education of Korea (BSRI-97-2401). I would like to thank 
S. Carlip, C. Lee, Q-H. Park, S. J. Hyun, and J.-H. Cho and 
S. W. Zoh for useful discussions and critical comments. 
I also would like to thank Professor R. Jackiw for introducing
me to some useful references.


\begin{thebibliography} {99}
\bibitem{comm1} In Witten's work
the $U(1)$ subgroup to be gauged is generated by $J_1$. But $J_1$ can be
transformed into $J_2$ by a global $SL(2,R)$ transformation. Thus,
the above gauged WZW model is quivalent to that of \cite{witten91}.
\end{thebibliography}
\end{document}